\documentclass{IOS-Book-Article}
\usepackage{threeparttable}
\usepackage{mathptmx}
\usepackage{amsmath}
\usepackage{tikz}
\usetikzlibrary{shapes,arrows}
\usepackage{multirow}
\usepackage{epstopdf}
\usepackage{enumerate}
\usepackage{tabularx}
\usepackage[T1]{fontenc}
\usepackage{babel}
\usepackage{amssymb}
\usepackage{bbm}
\usepackage{apacite}
\begin{document}
\begin{frontmatter}              

\title{The Application of Accumulation Test in Peaks Over Threshold Modeling with Norwegian Fire Insurance Data}
\runningtitle{The Application of Accumulation Test in Peaks Over Threshold Modeling with Norwegian Fire Insurance Data}

\author[A]{\fnms{Bowen} \snm{Liu}%
\thanks{Corresponding Author: Bowen Liu; E-mail:bowen.liu@umkc.edu}},
\author[B]{\fnms{Malwane M.A.} \snm{Ananda}}

\runningauthor{Bowen Liu}
\address[A]{Division of Computing, Analytics, and Mathematics, University of Missouri-Kansas City}
\address[B]{Department of Mathematical Sciences, University of Nevada, Las Vegas}

\begin{abstract}
Modeling excess remains to be an important topic in insurance data modeling. Among the alternatives to modeling excess, the Peaks Over Threshold (POT) framework with Generalized Pareto distribution (GPD) is regarded as an efficient approach due to its flexibility. However, selecting an appropriate threshold for such a framework is a major difficulty. To address such difficulty, we applied several accumulation tests along with Anderson-Darling test to determine an optimal threshold. Limited simulations were conducted to assess the performance of accumulation tests. Based on the selected thresholds, the fitted GPD with the estimated quantiles can be found. We applied the procedure to the well-known Norwegian Fire Insurance data and constructed the confidence intervals for the Value-at-Risks (VaR). The accumulation test approach provides satisfactory performance in modeling the high quantiles of Norwegian Fire Insurance data compared to the previous graphical methods.
\end{abstract}

\begin{keyword}
Generalized Pareto Distribution\sep Sequential Selection Procedure\sep Peaks Over Threshold Modeling\sep Tail Risk Estimation\sep Accumulation Tests
\end{keyword}
\end{frontmatter}

\thispagestyle{empty}
\pagestyle{empty}

\section{Introduction}

The selection of a proper risk modeling approach is a major topic in the insurance industry. Among all existing approaches, parametric modeling is preferred due to its flexibility and interpretability. Within parametric methods, the Peaks Over Threshold (POT) approach with Generalized Pareto Distribution (GPD) remains popular due to the support of the Pickands–Balkema–De Haan theorem \cite{mcneil_estimating_1997}. However, the performance of such approach is mainly determined by the selection of thresholds. If the threshold is chosen to be too low, the excess distribution might not be well approximated by the GPD; If the threshold is chosen to be too high, number of excesses would not be enough to establish a useful model.  

\par To identify an appropriate threshold in GPD modeling, many different methods have been proposed in the literature \cite[see]{langousis_threshold_2016,scarrot}. Gertensgarbe-Werner (GW) plot was first developed to identify the change point that separates non-extreme and extreme parts of the data \cite{gert_1989}. However, the uniform assumption on the data of this method is an essential limitation. Therefore, this method should be avoided in real applications \cite{langousis_threshold_2016}. Mean Residual Life (MRL) plot was also widely used to determine the GPD thresholds \cite{davison_models_1990,lang_towards_1999,coles_threshold_2001}. This method aims to identify the threshold by looking for the linear trend of the candidate thresholds on the GPD parameters. However, the selection of threshold is usually done by a visual inspection, which is subjective. Compared to the methods mentioned above, threshold selection procedures with Goodness-of-Fit (GoF) tests could be interpreted with better support from statistical theory.      

\par The objective of the GPD threshold selection procedures based on the GoF tests is to identify the lowest threshold such that the exceedances above the threshold fit a GPD well. For a single fixed threshold, the GoF test could be easily carried out \cite{choulakian_goodness--fit_2001}. However, a candidate set of possible thresholds are usually involved while finding the 'optimal' threshold. Thus, error control is necessary since the candidate thresholds are tested simultaneously, which generates a multiple-testing problem. Due to the natural ordering of the candidate thresholds, the hypotheses for the GoF tests should also be ordered. Therefore, the False Discovery Rate (FDR) control procedures such as the Benjamini-Hochberg \cite{benjamini_controlling_1995}  or Benjamini-Yekutieli \cite{benjamini_control_2001}  cannot be used without modifications. To adapt the FDR control procedures to the ordered hypotheses testing structure, \cite{gsell_sequential_2016} proposed a procedure named ForwardStop. ForwardStop provides a way to choose a stopping point for ordered hypotheses such that the hypotheses occur before the stopping point can be rejected with adequate error control. \cite{bader_automated_2018} then adapted this procedure to the selection of GDP thresholds with the application to the rainfall data.  
\par The accumulation tests \cite{li_accumulation_2017} could be seen as a generalization of the ForwardStop procedure. \cite{li_accumulation_2017} showed that, any functions that satisfy certain regularity conditions could be utilized in the ordered hypotheses testing procedure with the FDR control at a desired level. In addition to ForwardStop, procedures like SeqStep \cite{barber_controlling_2015} or HingeExp \cite{li_accumulation_2017} could also control FDR under different scenarios. 
\par In this paper, we attempted to adapt both SeqStep and HingExp procedures to the GPD threshold selection problems, with an application to the Norwegian fire insurance data. The rest of the paper is organized as follows: In section 2, an introduction of GPD model and POT method was given, with a brief introduction of accumulation tests for sequential hypotheses testing. Section 3 provides simulations to assess the performance of different accumulation tests in selecting GPD thresholds. The real data analysis of the Norwegian fire insurance data set is given in Section 4. The final section concludes with a discussion and our thoughts about the future directions.

\section{Methodology}

\subsection{POT Modeling and Threshold Selection Procedure}
A common practice in extreme value modeling is to apply the Peaks Over Threshold (POT) approach to the exceedances beyond a certain threshold $\mu$. This approach is built upon the second theorem of extreme value theory (the Pickands-Balkema-De Haan theorem). Under appropriate regularity conditions, the conditional excess function $F_u$ of a sequence of i.i.d. random variable $\{X_1, X_2, ...\}$ can be well approximated by a Generalized Pareto Distribution with location parameter $u$. A Generalized Pareto distribution is defined with the following cumulative distribution function (CDF):
\begin{equation}
    F(x|\gamma,\sigma,\mu)  = \begin{cases}
   1-(1+\frac{\gamma(x-\mu)}{\sigma})^{-\frac{1}{\gamma}} &  \gamma \neq  0 \\
     1-e^{-\frac{x-\mu}{\sigma}} &  \gamma = 0 \\
     \end{cases}
\end{equation}
where $x>\mu$. When $\gamma = 0$, the GPD distribution is defined as an exponential distribution with parameter $\sigma$. 
\par The choice of $\mu$ is important in establishing a valuable model when applying the POT method. By the second theorem of the extreme value theory, the exceedances over the sufficiently high threshold $\mu$ follow a GPD approximately. However, if the threshold is chosen to be too high, the corresponding fitted GPD model would have a poor ability to describe the data distribution. 
\par \indent Let $X_1, X_2, ..., X_n$ be a random sample of size $n$. Assume this random sample is taken such that the regularity conditions for the Pickand-Balkema-De Haan theorem hold. Then, for an appropriate threshold $\mu$, the exceedances $Y_i = X_i - \mu$ approximately follows a GPD. Thus, the objective is to identify the lowest possible $\mu$ such that the exceedances approximately follow GPD. 
\subsection{MLE of $\sigma$ and $\gamma$ when $\mu$ is known}
Suppose the threshold $\mu$ is known and both $\sigma$ and $\gamma$ are unknown. Suppose $\mu$ is fixed, let $Y_i = X_i -\mu$ for $i= 1,2,...,n$. The log-likelihood of $Y_1, Y_2, ..., Y_n$ is as follows: 
\begin{equation}
l(\sigma, \gamma| Y_1,Y_2, ..., Y_n) = -n \text{log}(\sigma)-(1+\frac{1}{\gamma})\sum_{i = 1}^{n} \text{log}(1+\frac{\gamma y_i}{\sigma})
\end{equation} 
Essentially, the maximum of the likelihood function can be obtained: 

\begin{equation*}
    \begin{cases}
     \frac{\partial l}{\partial \sigma} = 0\\
    \frac{\partial l}{\partial \gamma} = 0 ,
    \end{cases}
\end{equation*}

Davison showed that the solution of the above system of equations can be reduced to finding the solution of equation with one variable, with the following steps \cite{davison_models_1990}:

\begin{enumerate}
    \item Let $\frac{\gamma}{\sigma} = \theta$. The log-likelihood function becomes:
    \begin{equation}
    \label{eq2}
    l(\theta, \gamma| Y_1,Y_2, ..., Y_n) = -n \text{log}(\gamma)+n \text{log}(\theta) -(1+\frac{1}{\gamma})\sum_{i = 1}^{n} \text{log}(1+ \theta y_i)
    \end{equation}
    \item For any fixed value of $\theta$, the ML estimate of the $\gamma$ given $\theta$ is determined by  $\frac{\partial l}{\partial \gamma} = 0$: 
    
    \begin{equation} 
    \label{eq3}\hat{\gamma} = \frac{  \sum_{i = 1}^{n} \text{log}(1+ \theta y_i)}{n}  
    \end{equation}. 
    \item Plug \ref{eq3} back to \ref{eq2}. The optimization becomes a one-dimensional search of $\theta$: 
    \begin{equation}
        \label{eq5}
            l(\theta) = -n - n \text{log}(-\frac{1}{n} \sum_{i=1}^{n} log(1+\theta y_i)) - \sum_{i=1}^{n}\text{log}(1+\theta y_i)
    \end{equation}
    \item Eventually, the estimates of $\gamma$ and $\sigma$ can be expressed as follows, given the estimate of $\theta$ has been determined by maximizing Equation \ref{eq5} numerically:
    \begin{equation}
    \begin{cases}
     \hat{\gamma} =  \frac{1}{n} \sum_{i=1}^{n} \text{log}(1-\hat{\theta}y_i)\\
    \hat{\sigma} = \frac{\hat{\gamma}}{\hat{\theta}} .
    \end{cases}
\end{equation}
\end{enumerate}

Grimshaw showed that the MLE for $\gamma$ and $\sigma$ might not exist in some situations \cite{grimshaw_computing_1993}. However, Choulakian pointed out that this is not likely to occur especially when the sample size is large enough \cite{choulakian_goodness--fit_2001}. Under most of the scenarios, the MLE of $\gamma$ and $\sigma$ can be obtained with the procedure introduced earlier in this subsection. 
\subsection{Goodness-of-Fit (GoF) Tests for GPD with Anderson-Darling Statistics}
In order to assess the performance of the GPD fit, several tests were developed generally based on the GoF tests such as Anderson-Darling (AD) test. The details for the AD test are provided as follows:
    Given a sample $z$ is arranged in increasing order as $z_{(1)}<z_{(2)}<...<z_{(n)}$, AD test statistic for the uniform distribution is defined as:
\begin{equation}
A^2 = -n - \frac{1}{n}\sum_{i = 1}^{n}(2i-1)(\text{log}(z_{(i)}) - \text{log}(1-z_{(n+1-i)}))
\end{equation}
Therefore, to use the AD test for assessing the GoF of any distribution with the CDF $F(x|\theta)$ properly defined, one could always use the probability integral transformation ($z_i = F(x_i|\hat{\theta})$) on the ordered sample first and apply the AD test for the uniform distribution. Notice $\hat{\theta}$ is the MLE of the parameter $\theta$ under $H_0$. \\

\subsection{Existing Method: Automated Threshold Selection Procedure}

\par The automated threshold selection for the GPD was developed by utilizing an ordered hypothesis testing procedure named ForwardStop \cite{gsell_sequential_2016, bader_automated_2018}. The selection procedure for the GPD could be summarized as follows: 
\begin{enumerate}
	\item Pick $l$ candidate thresholds as $\mu_1<\mu_2<...<\mu_l$. 
	\item For each $\mu_j$ ($j \in \{1,2,...,l\}$), shift the data by applying the transformation $Y_i = X_i - \mu_j$ ($i \in \{1,2,..., n \}$). $n$ is the sample size. After the shift, only the non-negative $Y_i$s are reserved since only the data values beyond the threshold $\mu_j$ are used in the fitting of a GPD model. Denote the non-negative $Y_i$s as follows:
 \begin{equation*}
     \mathbf{Y}_{(j)} = \{Y_i | Y_i \geq 0 \}
 \end{equation*}
 Therefore, with $\mathbf{Y}_{(j)}$ determined by each $\mu_j$, The corresponding MLE of $\gamma$ and $\sigma$ ($\gamma_{j}$ and $\sigma_{j}$) can be obtained by using the method introduced in section 2.2,  
	
	\item Then, for each $\mu_j$, the null hypothesis:
 \begin{center}
 $H_j: \mathbf{Y}_{(j)}$ is drawn from a GPD distribution, 
 \end{center}
 could be examined by using GoF tests. $l$ different p-values are obtained correspondingly. 
	
    \item The largest index for the candidate thresholds to be rejected is determined by applying the ForwardStop procedure \cite{gsell_sequential_2016} as follows:  
    \begin{equation}
    \hat{k} = \text{argmax}_{k \in \{1,...,l\}}\{ \frac{1}{k}\sum_{i =1}^{k}-\text{log}(1-p_i)  \leq \alpha\}, 
    \end{equation} where $\alpha$ is pre-specified before the selection procedure. $\mu_{\hat{k}+1}$ is then the optimal threshold. 

\end{enumerate} 

The above procedure could produce good estimates for the thresholds of GPD if the candidates are chosen properly. For instance, when modeling the rainfall data,  \cite{bader_automated_2018} selected the thresholds within the $70^{\text{th}}$ and the $98^{\text{th}}$ percentile of the rainfall data and obtained a good geographical explanation based on the results obtained from the automated threshold selection procedure. 

\subsection{Accumulation Tests}
Li and Barber \cite{li_accumulation_2017} later generalized the results from different ordered hypothesis testing procedures \cite{gsell_sequential_2016, barber_controlling_2015} including ForwardStop. A family of accumulation tests were developed to choose a cutoff point $k$ such that first $k$ hypotheses are rejected, with a modified false discovery rate (FDR) being controlled. Assume $n$ hypotheses are ordered sequentially as $H_1, H_2, ..., H_n$ with corresponding $p$-values $p_1, p_2, ..., p_n$. An accumulation function is defined as an integrable nondecreasing function $h$ that maps $[0,1]$ to $[0, \infty]$. Then an accumulation test associated with the function $h$ chooses the cutoff $\hat{k}$ as: 
\begin{equation*}
    \hat{k} = \text{argmax}_{k \in {1,...,l}}\{ \frac{1}{k}\sum_{i =1}^{k}h(p_i) \leq \alpha\}, 
\end{equation*} where the hypotheses $H_1,H_2, ..., H_{\hat{k}}$ can be rejected at the pre-specified FDR level $\alpha$. Therefore, ForwardStop is essentially a special case of the accumulation test procedure when $h(p) = -log(1-p)$. 

\tikzstyle{decision} = [diamond, draw, fill=blue!20, 
text width= 5em, text badly centered, node distance=3cm, inner sep=0pt]
\tikzstyle{block} = [rectangle, draw, fill=blue!20, 
text width=10em, text centered, rounded corners, minimum height=4em]
\tikzstyle{line} = [draw, -latex']
\tikzstyle{cloud} = [draw, ellipse,fill=red!20, node distance=3cm,
minimum height=2em]
\begin{figure}[]
	\centering
\begin{tikzpicture}[node distance = 2.5cm, auto]
\node [block] (init) {Initialize candidate thresholds ($\mu_1, \mu_2, ..., \mu_{m}$) in ascending order};
\node [block, below of=init] (identify) {Find p-values for all candidates correspondingly. ($p_1, p_2, ..., p_m$)};
\node [block, below of=identify] (evaluate) {Calculate the accumulation test statistics ($\frac{1}{k}\sum_{i=1}^{k}h(p_i)$ for all $k \in {1,2,...,m}$) };
\node [decision, below of=evaluate] (decide) {Do any of the test statistics exceed $\alpha$?};
\node [block, left of=decide, node distance=5cm] (null) {No threshold can be identified within the candidates};
\node [block, right of=decide, node distance=5cm] (stop) {Identify $\hat{k}$ based on (3) and $\mu_{\hat{k}+1}$ is chosen as the threshold.};
\path [line] (init) -- (identify);
\path [line] (identify) -- (evaluate);
\path [line] (evaluate) -- (decide);
\path [line] (decide) -- node {no} (null);
\path [line] (decide) -- node {yes} (stop);

\end{tikzpicture}
\vspace*{0.3cm}

\caption{The Flow Chart of The Accumulation Test Threshold Selection Framework}
\label{fig:flow}
\end{figure}
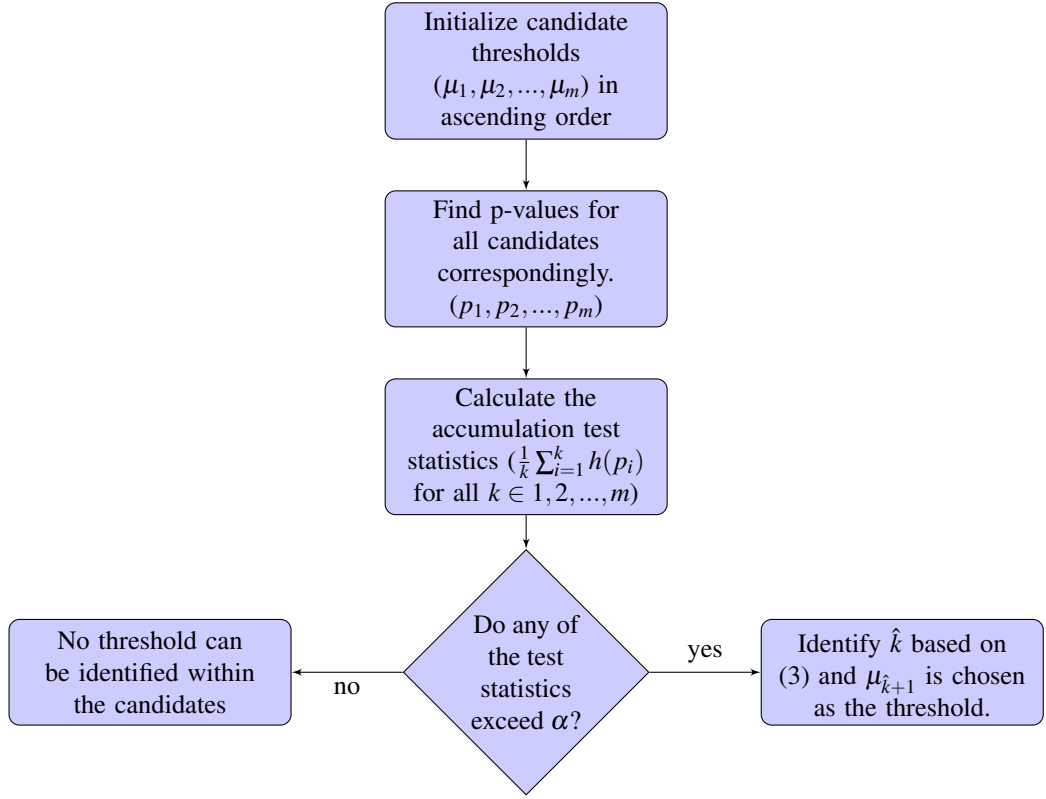
\par Therefore, if we use any function $h$ that satisfies the conditions for the accumulation functions, a generalized GPD threshold selection procedure can be implemented along with the automated threshold selection procedure developed by Bader \cite{bader_automated_2018}. The flow chart for the generalized GPD threshold selection procedure is summarized in Figure \ref{fig:flow}. 

\subsection{VaR and the interval estimation of VaR for GPD}
The estimation of Value-at-Risk (VaR) is important for the insurance data modeling. For a loss random variable, VaR at the level of $p$ is defined as:
\begin{equation*}
	P(X<VaR_{p}(X)) = p,
\end{equation*}
In the insurance industry, VaR is the amount of capital that one insurance company needs to maintain to prevent the company from bankruptcy due to very extreme claims. Given that a random variable $X$ follows a GPD given CDF in (1), the VaR of $X$ could be easily derived in the following form:
\begin{equation}
    VaR_p(X) = \mu+\frac{\sigma}{\gamma}[(1-p)^{-\gamma}-1]
\end{equation}
Therefore, for given $\mu$, $\hat{\sigma}$ and $\hat{\gamma}$, the ML estimate of $VaR_p(X)$ is:
\begin{equation*}
    \widehat{VaR_p(X)} = \hat{\mu}+\frac{\hat{\sigma}}{\hat{\gamma}}[(1-p)^{-\hat{\gamma}}-1]
\end{equation*}
\par As we explained in section 2.2, the MLE of $\sigma$ and $\gamma$ could be obtained numerically for a given $\mu$. \cite{smith_threshold_1984} proved that when $\sigma > -\frac{1}{2}$, the MLE of $\sigma$ and $\gamma$ are asymptotically normal and efficient:
\begin{equation*}
    \begin{bmatrix}
        \hat{\sigma} \\
         \hat{\gamma} \\
    \end{bmatrix} \sim \text{AN} 
    \begin{pmatrix}
     \begin{bmatrix}
         \sigma \\
         \gamma \\ 
     \end{bmatrix},
     \Sigma
     
    \end{pmatrix}, \end{equation*} where $\Sigma =  n^{-1} \begin{bmatrix}
         2 \sigma^2 (1+\gamma) & \sigma (1+\gamma) \\
         \sigma (1+\gamma) & (1+\gamma)^2\\ 
     \end{bmatrix}$.
 With the MLE of $\sigma$ and $\gamma$, the asymptotic variance of $\widehat{VaR_p(X)}$ could be derived from the multivariate Delta method. Consider $VaR_p(X) = g(\sigma,\gamma|\mu) = \mu+\frac{\sigma}{\gamma}[(1-p)^{-\gamma}-1]$, the gradient $\nabla g = [\frac{(1-p)^{-\gamma}-1}{\gamma}, \frac{\sigma (1-p)^{-\gamma}[(1-p)^{\gamma}-\gamma \text{log}(1-p)-1]}{\sigma^2}]^T$. By Delta method, $\widehat{VaR_p(X)}$ has the asymptotic distribution $N(VaR_p(x), \omega^2)$, where
 \begin{align}
   \omega^2 = \nabla g^T \Sigma \nabla g & = 2 {\frac{(1-p)^{-\gamma}-1}{\gamma^2}} \sigma^2 (1+\gamma) +{\frac{\sigma (1-p)^{-\gamma}[(1-p)^{\gamma}-\gamma \text{log}(1-p)-1]}{\sigma^2}}^2 (1+\gamma)^2 \\ &  +2 \frac{(1-p)^{-\gamma}-1}{\gamma} \frac{\sigma (1-p)^{-\gamma}[(1-p)^{\gamma}-\gamma \text{log}(1-p)-1]}{\sigma^2} \sigma (1+\gamma)
 \end{align}. 
By utilizing Delta method, the asymptotic $100(1-\alpha) \%$ CI for $VaR(X)$ is is given as:
\begin{equation}
    \widehat{VaR_p(X)} \pm z_{\alpha/2} \hat{\omega},
\end{equation}
where $z_{\alpha}$ represents the $\alpha^{\text{th}}$ quantile of a standard Gaussian random variable.

\section{Simulation}

\subsection{Simulation Settings}
To assess the performance of sequential testing procedures, we conducted limited simulations. Three different accumulation tests were assessed for our simulations. The accumulation functions of these three accumulation tests are provided as follows:
\begin{itemize}
    \item \textit{ForwardStop:} $h(p) = -\text{log}(1-p)$
    \item \textit{SeqStep:} $h(p) = C  \cdot \mathbbm{1}_{ \{p>1-\frac{1}{C}\}}$
    \item \textit{HingeExp:} $h(p) = C  \cdot \mathbbm{1}_{ \{p>1-\frac{1}{C}\}} \text{log}(C(1-p))$
\end{itemize}
For the value of $C$ involved in SeqStep and HingExp, we selected $C=2$ as Li and Barber \cite{li_accumulation_2017} recommended in their work of accumulation tests. For the significance threshold $\alpha$, we selected $\alpha = 0.01$ and $\alpha = 0.05$ in the simulations.

\subsection{Scenarios of composite density with GPD tails}
To demonstrate the ability of accumulation tests in selecting the thresholds for a GPD distribution in a POT model, we generated samples from four parametric composite distributions. The composite distributions were widely used in the modeling of insurance claim sizes \cite{ananda2005, scollnik_composite_2007,Scollnik2012ModelingWW,brazauskas_modeling_2016, grun2019, liu_analyzing_2022, liu_generalized_2022, mutali_composite_2020, deng_bayesian_2019, ig_pareto, exp_pareto, calderin-ojeda_note_2018}. The details of the simulation scenarios are listed in Table 1. For each scenario, $r = 1,000$ replicates were generated. 

The mean and the RMSE were used to assess the performances of three tests under all scenarios. The formula for RMSE is provided as follows:
\begin{equation*}
    RMSE = \sqrt{\frac{\sum_{i = 1}^{r}(\hat{\mu}_i - \mu)^2}{r}},
\end{equation*}
where $\hat{\mu}_i$ denotes the chosen threshold for the $i$-th replicate, $\mu$ stands for the true threshold under each scenario, and $r$ is the number of replicates.

\subsection{Simulation Results}

The simulation results are presented in Table 1 ($\alpha = 0.01$) and Table 2 ($\alpha = 0.05$). For all scenarios, ForwardStop showed great performance, while SeqStep and HingeExp provided unsatisfactory performance. Notice for all sequential testing procedures, the RMSE decrease as the sample size increases. 

\begin{table}[]
\label{table:table_sim_2}
\centering
	\caption{Simulation Results for Three Different Tests Under Different Simulation Scenarios ($\alpha = 0.01$)}
		\footnotesize
\begin{tabular}{lllllllllll}
\hline
\multirow{2}{*}{\textbf{Scenario}}                                                                                                     & \multirow{2}{*}{\textbf{\begin{tabular}[c]{@{}l@{}}True \\ Threshold\end{tabular}}} & \multirow{2}{*}{\textbf{\begin{tabular}[c]{@{}l@{}}Head \\ Weight\end{tabular}}} & \multirow{2}{*}{\textbf{\begin{tabular}[c]{@{}l@{}}Tail \\ Weight\end{tabular}}} & \multirow{2}{*}{\textbf{\begin{tabular}[c]{@{}l@{}}Sample\\ Size\end{tabular}}} & \multicolumn{2}{l}{\textbf{\begin{tabular}[c]{@{}l@{}}Forward\\ Stop\end{tabular}}} & \multicolumn{2}{l}{\textbf{\begin{tabular}[c]{@{}l@{}}SeqStep\\ (C=2)\end{tabular}}} & \multicolumn{2}{l}{\textbf{\begin{tabular}[c]{@{}l@{}}HingeExp\\ (C=2)\end{tabular}}} \\ \cline{6-11} 
                                                                                                                                       &                                                                                     &                                                                                  &                                                                                  &                                                                                 & \textbf{Mean}                            & \textbf{RMSE}                            & \textbf{Mean}                             & \textbf{RMSE}                            & \textbf{Mean}                             & \textbf{RMSE}                             \\ \hline
\multirow{3}{*}{GPD$(\mu = 0.5, \sigma =1, \gamma = 1)$}                                                                                    & \multirow{3}{*}{0.5}                                                                & \multirow{3}{*}{-}                                                               & \multirow{3}{*}{-}                                                               &  n = 100                                                                         &                     0.517                     &     0.047                                     &           2.223                                &                    1.042                      &       5.273                                    &              1.692                                                                                              \\
                                                                                                                                       &                                                                                     &                                                                                  &                                                                                  & n = 200                                                                         &            0.514                              &        0.041                                  &                        2.012                   &                  0.876                        &                  4.973                         &      1.474                                     \\
                                                                                                                                       &                                                                                     &                                                                                  &                                                                                  & n = 500                                                                         &    0.508                                      &         0.007                                 &                         1.842                  &             0.516                             &           4.549                                &      1.025                                     \\ \hline
\multirow{3}{*}{GPD$(\mu = 2, \sigma =1, \gamma = 1)$}                                                                                      & \multirow{3}{*}{0.5}                                                                & \multirow{3}{*}{-}                                                               & \multirow{3}{*}{-}                                                               & n = 100                                                                         &                            2.020              &     0.013                                     &                       3.333                    &                      0.665                    &      8.664                                     &                     2.118                      \\
                                                                                                                                       &                                                                                     &                                                                                  &                                                                                  & n = 200                                                                         &        2.016                                  &        0.009                                  &                        3.307                   &           0.487                               &                8.177                           &    1.764                                       \\
                                                                                                                                       &                                                                                     &                                                                                  &                                                                                  & n = 500                                                                         &   2.009                                       &        0.003                                  &                            3.255               &               0.182                           &          7.912                                 &      0.781                                     \\ \hline
\multirow{3}{*}{\begin{tabular}[c]{@{}l@{}}lognormal$(\mu = 2,\sigma = 0.5)$,\\ GPD$(\mu = 2, \sigma = 0.8, \gamma=0.8)$\end{tabular}} & \multirow{3}{*}{2}                                                                  & \multirow{3}{*}{0.3}                                                             & \multirow{3}{*}{0.7}                                                             & n = 100                                                                         &  1.941                                        &     0.080                                     &      5.575                                     &     1.133                                     &     9.497                                      &         1.549                                  \\
                                                                                                                                       &                                                                                     &                                                                                  &                                                                                  & n = 200                                                                         &        1.953                                  &        0.069                                  &                           4.892                &            0.984                              &           8.768                                &     1.376                                      \\
                                                                                                                                       &                                                                                     &                                                                                  &                                                                                  & n = 500                                                                         &     1.981                                     &       0.091                                   &                    4.137                       &                 0.582                         &                     7.642                      &    1.058                                       \\ \hline
\multirow{3}{*}{\begin{tabular}[c]{@{}l@{}}lognormal$(\mu=1,\sigma = 1)$, \\ GPD$(\mu = 1, \sigma = 1.3, \gamma = 1.5)$\end{tabular}}  & \multirow{3}{*}{1}                                                                  & \multirow{3}{*}{0.2}                                                             & \multirow{3}{*}{0.8}                                                             & n = 100                                                                         &  0.896                                  &      0.113                                  &                      9.899             &                                  4.164      &               43.735                   &           23.396                          \\
                                                                                                                                       &                                                                                     &                                                                                  &                                                                      & n = 200                                                                         &           0.914                               &           0.085                               &                    9.041                       &                     3.569                     &                  35.679                         &                     17.773                      \\
                                                                                                                                       &                                                                                     &                                                                                  &                                                                                  & n = 500                                                                         &           0.946                               &        0.038                                  &                           8.152                &             2.146                             &            29.177                               &                      12.380                     \\ \hline
\end{tabular}
\end{table}

\begin{table}[]
\label{table:table_sim}
\centering
	\caption{Simulation Results for Three Different Tests Under Different Simulation Scenarios ($\alpha = 0.05$)}
		\footnotesize
\begin{tabular}{lllllllllll}
\hline
\multirow{2}{*}{\textbf{Scenario}}                                                                                                     & \multirow{2}{*}{\textbf{\begin{tabular}[c]{@{}l@{}}True \\ Threshold\end{tabular}}} & \multirow{2}{*}{\textbf{\begin{tabular}[c]{@{}l@{}}Head \\ Weight\end{tabular}}} & \multirow{2}{*}{\textbf{\begin{tabular}[c]{@{}l@{}}Tail \\ Weight\end{tabular}}} & \multirow{2}{*}{\textbf{\begin{tabular}[c]{@{}l@{}}Sample\\ Size\end{tabular}}} & \multicolumn{2}{l}{\textbf{\begin{tabular}[c]{@{}l@{}}Forward\\ Stop\end{tabular}}} & \multicolumn{2}{l}{\textbf{\begin{tabular}[c]{@{}l@{}}SeqStep\\ (C=2)\end{tabular}}} & \multicolumn{2}{l}{\textbf{\begin{tabular}[c]{@{}l@{}}HingeExp\\ (C=2)\end{tabular}}} \\ \cline{6-11} 
                                                                                                                                       &                                                                                     &                                                                                  &                                                                                  &                                                                                 & \textbf{Mean}                            & \textbf{RMSE}                            & \textbf{Mean}                             & \textbf{RMSE}                            & \textbf{Mean}                             & \textbf{RMSE}                             \\ \hline
\multirow{3}{*}{GPD$(\mu = 0.5, \sigma =1, \gamma = 1)$}                                                                                    & \multirow{3}{*}{0.5}                                                                & \multirow{3}{*}{-}                                                               & \multirow{3}{*}{-}                                                               

&  n = 100                               &                     0.662                    &     0.056                                     &           3.126                                &                    1.242                      &       7.134                                    &              1.719                                                                                              \\
                                                                                                                                       &                                                                                     &                                                                                  &                                                                                  & n = 200                                                                         &            0.597                              &        0.050                                  &                        2.124                   &                  0.772                        &                  6.171                         &      1.351                                     \\
                                                                                                                                       &                                                                                     &                                                                                  &                                                                                  & n = 500                                                                         &    0.568                                      &         0.011                                 &                         1.939                  &             0.572                             &           5.813                                &      0.997                                     \\ \hline
\multirow{3}{*}{GPD$(\mu = 2, \sigma =1, \gamma = 1)$}                                                                                      & \multirow{3}{*}{0.5}                                                                & \multirow{3}{*}{-}                                                               & \multirow{3}{*}{-}                                                               & n = 100                                                                         &                            2.514              &     0.021                                &                       3.992                    &                      0.785                    &      12.178                                     &                     2.375                   \\
                                                                                                                                       &                                                                                     &                                                                                  &                                                                                  & n = 200                                                                         &        2.423                                  &        0.015                                  &                        3.687                   &           0.512                               &                10.045                           &    1.891                                       \\
                                                                                                                                       &                                                                                     &                                                                                  &                                                                                  & n = 500                                                                         &   2.291                                       &        0.011                                  &                            3.491               &               0.195                           &          9.912                                 &      0.748                                     \\ \hline
\multirow{3}{*}{\begin{tabular}[c]{@{}l@{}}lognormal$(\mu = 2,\sigma = 0.5)$,\\ GPD$(\mu = 2, \sigma = 0.8, \gamma=0.8)$\end{tabular}} & \multirow{3}{*}{2}                                                                  & \multirow{3}{*}{0.3}                                                             & \multirow{3}{*}{0.7}                                                             & n = 100                                                                         &  2.242                                        &     0.125                                     &      6.914                                     &     1.253                                     &     11.497                                      &         1.923                                  \\
                                                                                                                                       &                                                                                     &                                                                                  &                                                                                  & n = 200                                                                         &        2.172                                  &        0.081                                  &                           5.997                &            0.991                              &           10.768                                &     1.568                                      \\
                                                                                                                                       &                                                                                     &                                                                                  &                                                                                  & n = 500                                                                         &     2.104                                     &       0.092                                   &                    5.237                       &                 0.693                         &                     9.645                      &    1.342                                       \\ \hline
\multirow{3}{*}{\begin{tabular}[c]{@{}l@{}}lognormal$(\mu=1,\sigma = 1)$, \\ GPD$(\mu = 1, \sigma = 1.3, \gamma = 1.5)$\end{tabular}}  & \multirow{3}{*}{1}                                                                  & \multirow{3}{*}{0.2}                                                             & \multirow{3}{*}{0.8}                                                             & n = 100                                                                         &  1.196                                  &      0.146                                  &                      10.891             &                                  3.928      &               33.614                   &           15.463                          \\
                                                                                                                                       &                                                                                     &                                                                                  &                                                                      & n = 200                                                                         &           1.131                               &           0.103                               &                    10.141                       &                     3.261                     &                  31.436                         &                     13.616 \\
                                                                                                                                       &                                                                                     &                                                                                  &                                                                                  & n = 500                                                                         &           1.074                               &        0.047                                  &                           9.352                &             2.646                             &            27.377                               &                      11.453                     \\ \hline
\end{tabular}
\end{table}

\section{Real Data Application: Threshold Selection}
In this section, the well-known Norwegian Fire Insurance data is used to assess the performance of the different methods of the GPD threshold selection. 

\subsection{Data}
\begin{figure}
	\centering
	\includegraphics[width=1\linewidth]{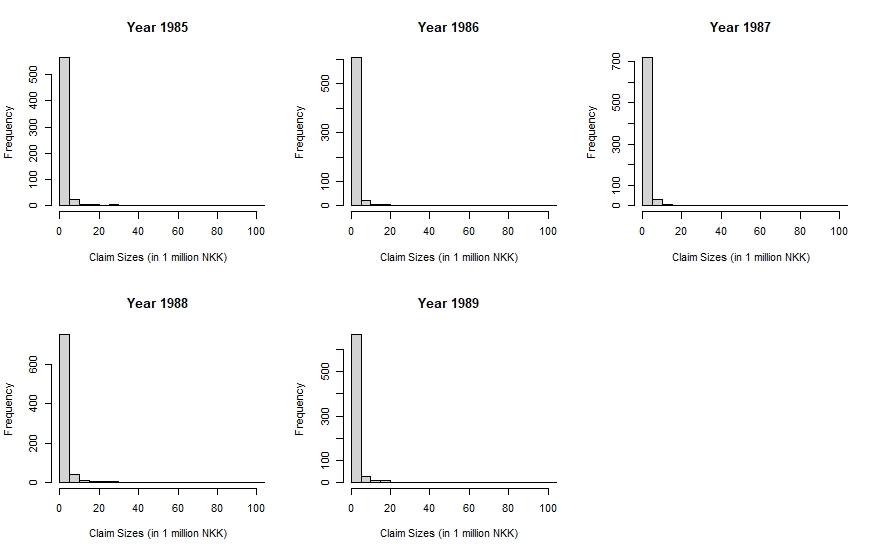}
	\vspace*{0.4cm}

	\caption{Histogram of Norwegian Fire Insurance Data (Year 1985-1989)}
	\label{fig:hist_1}
\end{figure}

To illustrate the performance of the accumulation tests on the GPD threshold selection problems, we utilized the Norwegian Fire Insurance Data. The data set contains the fire insurance claims from a Norwegian company from year 1972 to 1992. The dataset is available via the R package 'ReIns' \cite{reins}. No information regarding inflation adjustments was provided so we chose the claims from year 1985 to 1989. Moreover, only the damages over 500,000 Norwegian Krones (NKK) are available. For analysis concern, we scaled the data by 1,000,000. The summary statistics of the claims from year 1985 to 1989 are presented in Table \ref{table:table1}. Histograms of the claim sizes from 1985 to 1989 are presented in Figure \ref{fig:hist_1}.

\begin{table*}[!htbp]
	\centering
	\caption{Summary Statistics for Norwegian Fire Insurance Claims (1985-1989)}
		\footnotesize

	\begin{tabularx}{\textwidth}{XXXXXXXX}
		\hline
		 \multirow{2}{4em}{Year}
		&\multirow{2}{6em}{Sample Size}  & \multirow{2}{6em}{Mean} & \multirow{2}{6em}{SD} & \multirow{2}{6em}{Q1}  & \multirow{2}{6em}{Q2} & \multirow{2}{6em}{Q3} & \multirow{2}{6em}{Maximum}
		\\
		
		\\
		
		\hline
		
		  \multirow{1}{4em}{1985}  &607 & 2.553 & 8.013 & 0.680 & 1.000 & 1.712 &135.080
		\\

			\multirow{1}{4em}{1986} 
		& 647 & 2.477 & 9.695 & 0.700 & 0.985 & 1.549 & 188.270
		\\
		
			\multirow{1}{4em}{1987} 
		& 767  & 2.057 & 3.644 & 0.755 & 1.138 & 1.853 & 44.926
		\\
		
			\multirow{1}{4em}{1988} 
		&  827 & 3.176 & 17.677 & 0.762 & 1.176 & 2.049 & 465.365
		\\
				\multirow{1}{4em}{1989} 
		& 718 & 2.400 & 7.094 & 0.751 & 1.183 & 1.996 & 145.156
		\\
		
		\hline

	\end{tabularx}
	\label{table:table1}
\end{table*}

\subsection{Selection of Accumulation Functions and Methods for Comparison}
The accumulation tests that we chose in the simulations were applied to the Norwegian fire insurance data set. In addition,
 we selected two most commonly-used methods (GW plot and MRL plot) for comparison purposes. We utilized the R packages "tea" and "eva" to construct the plots and select the GPD thresholds. 

\subsection{Result: Threshold Selection}

\begin{figure}[]
	\centering
	\includegraphics[width=1\linewidth]{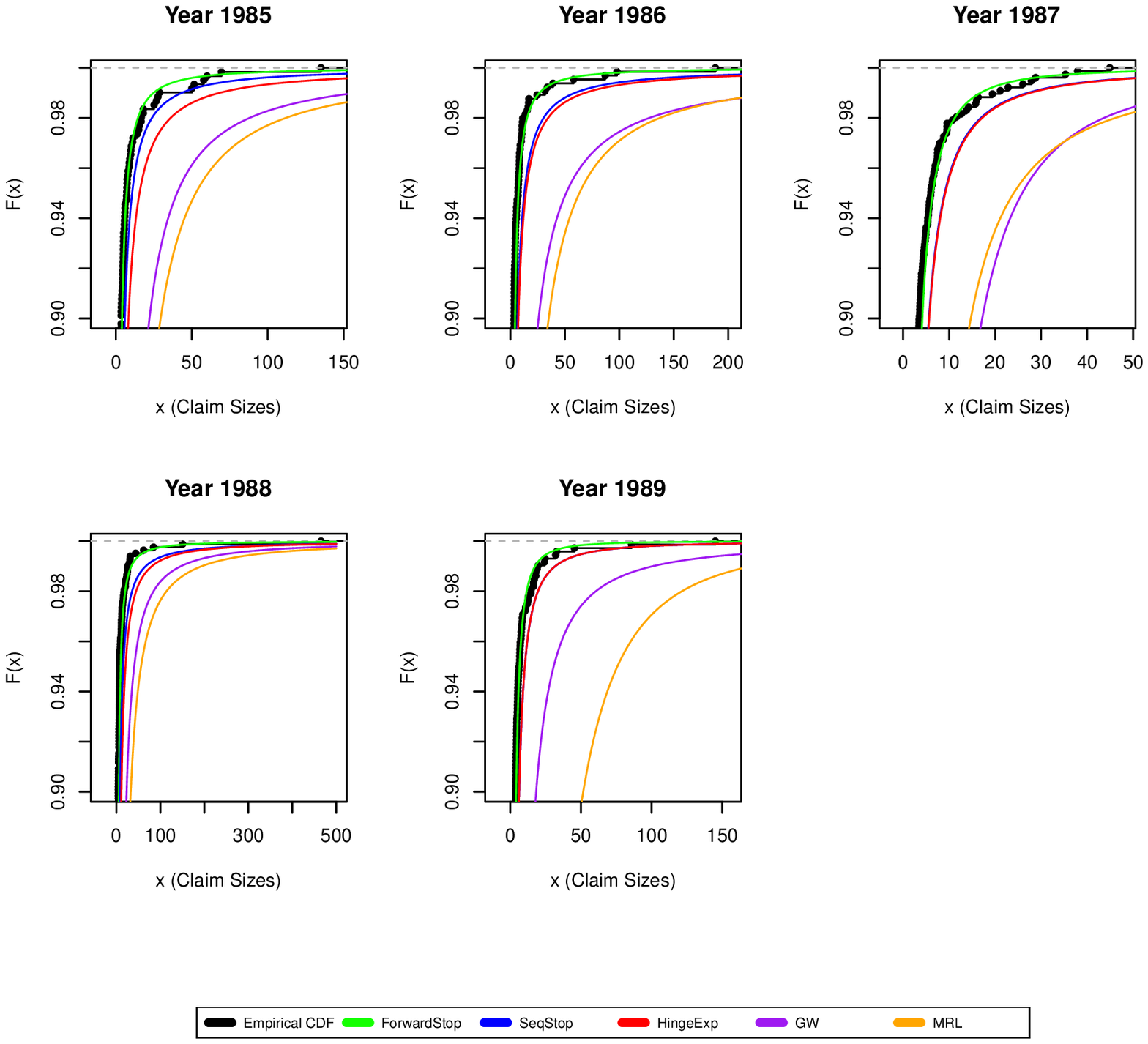}

	\caption{Comparison of the CDFs using different GPD modeling methods for Norwegian Fire Insurance Data (Year 1985-1989) ($\alpha = 0.01$ for the accumulation tests)}
	\label{fig:ecdf1}
\end{figure}

\begin{figure}[]
	\centering
	\includegraphics[width=1\linewidth]{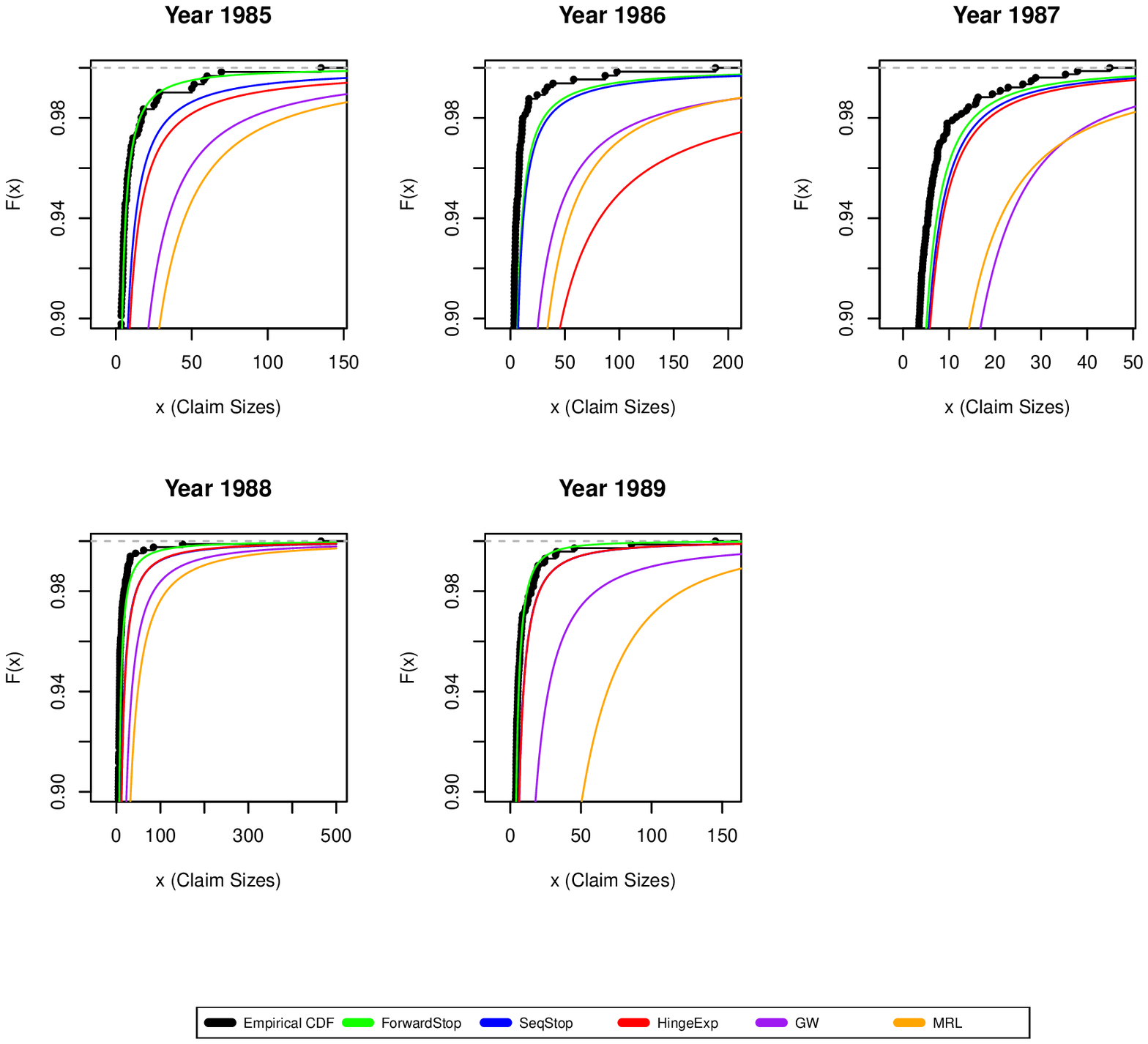}

	\caption{Comparison of the CDFs using different GPD modeling methods for Norwegian Fire Insurance Data (Year 1985-1989) ($\alpha = 0.05$ for the accumulation tests)}
	\label{fig:ecdf2}
\end{figure}

Table \ref{fig:table2} presents the selection of the GPD thresholds $\mu$, the corresponding MLE of the scale parameter $\sigma$ and the shape parameter $\gamma$ for the Norwegian Fire Insurance Data with the GW plot and the MRL plot; Table \ref{fig:table3}, \ref{fig:table4} and \ref{fig:table5} demonstrate the selection of the GPD thresholds $\mu$, the corresponding MLE of the scale parameter $\sigma$ and the shape parameter $\gamma$ for the Norwegian Fire Insurance Data with the ForwardStop selection, the SeqStep selection and HingeExp selection. Notice the GW plot and the MRL plot methods provided higher estimates for $\mu$ compared to the selection methods based on the accumulation tests. With these estimates, we were able to create the estimated CDFs of the GPD for the Norwegian Fire Insurance Data, with the different threshold selection methods. Comparisons of the estimated CDFs and the empirical CDFs for the Norwegian Fire Insurance Data are demonstrated in Figures \ref{fig:ecdf1} and \ref{fig:ecdf2}. 
\par We noticed that the estimated CDFs using the threshold selection with accumulation tests are closer to the empirical CDFs than the graphical methods over all the years we selected from the Norwegian Fire Insurance Data. This implies that the accumulation tests have a stronger ability to detect the appropriate thresholds compared to the graphical methods, with the GPD assumptions. Among three accumulation tests, the ForwardStop selection provided the closest fits to the empirical CDFs. Notice if we compare Tables \ref{fig:table3}, \ref{fig:table4} and \ref{fig:table5}, the selected thresholds with ForwardStop are significantly lower compared to the selected thresholds with HingeExp or SeqStep. This could be relevant since, as mentioned in Section 3.1, only claims with very high values (over 500,000 NKKs) are available. Given the data is obtained in this way, it is likely that the distribution of most of the claims over 500,000 NKKs could be modeled appropriately using GPD, based on the second theorem of extreme theory. Since the ForwardStop selection provided the lowest thresholds among all of the methods mentioned in this study, it obviously includes more claims for GPD modeling in comparison to other methods that chose higher thresholds. Thus, it is reasonable that the ForwardStop selection provided the closest fits to empirical CDFs comapred to all other methods.  
\par It is also notable that the choices of the false discovery proportion $\alpha$ have impacts on the selection of the thresholds. With the accumulation tests, $\alpha = 0.01$ (Figure \ref{fig:ecdf1}) provides better fits to the emiprical CDF compared to $\alpha = 0.05$ (Figure \ref{fig:ecdf2}), especially for year 1986 and 1987. A possible explanation is that, when $\alpha$ is set to be low, the thresholds would be detected earlier based on what was described in section 2.5. Therefore, with the nature of the Norwegian Fire Insurance Data being truncated at 0.5 million NKK, $\alpha = 0.01$ apparently produces the thresholds closer to the truncated point, and hence leads to a closer fit to the data in comparison with $\alpha = 0.05$.  
\begin{table*}[!htbp]
	\centering
	\caption{Selection of Thresholds ($\mu$) with MLEs of scale ($\sigma$) and shape ($\gamma$) parameters for Norwegian Fire Insurance Claims (1985-1989) Using GW Plot and MRL Plot}
	\footnotesize
	\begin{tabularx}{\textwidth}{XXXXX}
		\hline
		\multirow{2}{4em}{Method} & \multirow{2}{4em}{Year}
		&\multirow{2}{6em}{Chosen Threshold ($\mu$)}  & \multirow{2}{6em}{Scale ($\sigma$)} & \multirow{2}{6em}{Shape ($\gamma$)}  
		\\
		
		\\
		
		\hline
		\multirow{5}{4em}{GW Plot}
		&  \multirow{1}{4em}{1985}  & 2.570 & 2.764 & 0.843 
		\\

		&	\multirow{1}{4em}{1986} 
		& 3.363 & 2.460 & 1.010 
		\\
		
		&	\multirow{1}{4em}{1987} 
		&3.499 &3.014 &0.536
		\\
		
		&		\multirow{1}{4em}{1988} 
		&  3.331 & 3.010 & 0.795 
		\\
		&		\multirow{1}{4em}{1989} 
		& 3.062 & 2.532 & 0.734
		\\
		
		\hline
		\multirow{5}{4em}{MRL Plot}
		&  \multirow{1}{4em}{1985}  &   3.500 & 3.854 & 0.810
		\\

		&	\multirow{1}{4em}{1986} 
		& 4.500 & 4.356 & 0.838
		\\
		
		&	\multirow{1}{4em}{1987} 
		&  3.000 & 1.998  & 0.716
		\\
		
		& 			\multirow{1}{4em}{1988} 
		&  5.000 & 4.388 & 0.766
		\\
		& 			\multirow{1}{4em}{1989} 
		&  8.500 & 10.797 & 0.441
		\\
		\hline

	\end{tabularx}
	\label{fig:table2}
\end{table*}

\begin{table*}[!htbp]
	\centering
	\caption{ForwardStop-Chosen thresholds ($\mu$) with MLEs of scale ($\sigma$) and shape ($\gamma$) parameters in the GPD Models for Norwegian Fire Insurance Claims (1985-1989)}
	\footnotesize
	\begin{tabularx}{\textwidth}{XXXXX}
		\hline
		\multirow{2}{4em}{$\alpha$} & \multirow{2}{4em}{Year}
		&\multirow{2}{6em}{Chosen Threshold ($\mu$)}  & \multirow{2}{6em}{Scale ($\sigma$)} & \multirow{2}{6em}{Shape ($\gamma$)}  
		\\
		
		\\
		
		\hline
		\multirow{5}{4em}{0.01}
		&  \multirow{1}{4em}{1985}  & 0.541 & 0.558 & 0.769
		\\

		&	\multirow{1}{4em}{1986} 
		& 0.677 & 0.539 & 0.798
		\\
		
		&	\multirow{1}{4em}{1987} 
		&0.660  & 0.757 & 0.554
		\\
		
		&		\multirow{1}{4em}{1988} 
		&  0.745 & 0.768 & 0.768
		\\
		&		\multirow{1}{4em}{1989} 
		& 0.531 & 0.764 & 0.567
		\\
		
		\hline
		\multirow{5}{4em}{0.05}
		&  \multirow{1}{4em}{1985}  &   0.577 & 0.524 & 0.823
		\\

		&	\multirow{1}{4em}{1986} 
		& 1.079 & 0.655 & 0.970
		\\
		
		&	\multirow{1}{4em}{1987} 
		&  0.974 & 0.770  & 0.662
		\\
		
		& 			\multirow{1}{4em}{1988} 
		&  0.973 & 0.871 & 0.811
		\\
		& 			\multirow{1}{4em}{1989} 
		&  0.555 & 0.751 & 0.584
		\\
		\hline

	\end{tabularx}
		\label{fig:table3}

\end{table*}

\begin{table*}[!htbp]
	\centering
	\caption{SeqStep (C=2) Chosen thresholds ($\mu$) with MLEs of scale ($\sigma$) and shape ($\gamma$) parameters in the GPD Models for Norwegian Fire Insurance Claims (1985-1989)}
	\footnotesize
	\begin{tabularx}{\textwidth}{XXXXX}
		\hline
		\multirow{2}{4em}{$\alpha$} & \multirow{2}{4em}{Year}
		&\multirow{2}{6em}{Chosen Threshold ($\mu$)}  & \multirow{2}{6em}{Scale ($\sigma$)} & \multirow{2}{6em}{Shape ($\gamma$)}  
		\\
		
		\\
		
		\hline
		\multirow{5}{4em}{0.01}
		&  \multirow{1}{4em}{1985}  & 0.862 & 0.677 & 0.874
		\\

		&	\multirow{1}{4em}{1986} 
		& 1.079 & 0.655 & 0.970
		\\
		
		&	\multirow{1}{4em}{1987} 
		&1.086 & 0.816 & 0.677
		\\
		
		&		\multirow{1}{4em}{1988} 
		&  1.248 & 0.918 & 0.909
		\\
		&		\multirow{1}{4em}{1989} 
		& 1.138 & 0.897 & 0.698
		\\
	
		\hline
		\multirow{8}{4em}{0.05}
		&  \multirow{1}{4em}{1985}  &   1.193 & 0.876 & 0.921
		\\

		&	\multirow{1}{4em}{1986} 
		& 1.152 & 0.688 & 0.997
		\\
		
		&	\multirow{1}{4em}{1987} 
		&  1.104 & 0.818  & 0.682
		\\
		
		& 			\multirow{1}{4em}{1988} 
		&  1.590 & 1.388 & 0.846
		\\
		& 			\multirow{1}{4em}{1989} 
		&  1.170 & 0.870 & 0.726
		\\
	
		\hline
		\\
	\end{tabularx}
		\label{fig:table4}

\end{table*}

\begin{table*}[!htb]
	\centering
	\caption{HingeExp (C=2) Chosen thresholds ($\mu$) with MLEs of scale ($\sigma$) and shape ($\gamma$) parameters in the GPD Models for Norwegian Fire Insurance Claims (1985-1989)}
	\footnotesize
	\begin{tabularx}{\textwidth}{XXXXX}
		\hline
		\multirow{2}{4em}{$\alpha$} & \multirow{2}{4em}{Year}
		&\multirow{2}{6em}{Chosen Threshold ($\mu$)}  & \multirow{2}{6em}{Scale ($\sigma$)} & \multirow{2}{6em}{Shape ($\gamma$)}  
		\\
		
		\\
		
		\hline
		\multirow{5}{4em}{0.01}
		&  \multirow{1}{4em}{1985}  & 1.214 & 0.885 & 0.926
		\\

		&	\multirow{1}{4em}{1986} 
		& 1.152 & 0.688 & 0.997
		\\
		
		&	\multirow{1}{4em}{1987} 
		&1.104  & 0.818 & 0.682
		\\
		
		&		\multirow{1}{4em}{1988} 
		&  1.590 & 1.388 & 0.846
		\\
		&		\multirow{1}{4em}{1989} 
		& 1.138 & 0.897 & 0.698
		\\
	
		\hline
		\multirow{5}{4em}{0.05}
		&  \multirow{1}{4em}{1985}  &   1.378 & 0.894 & 1.004
		\\

		&	\multirow{1}{4em}{1986} 
		& 6.972 & 3.524 & 1.152
		\\
		
		&	\multirow{1}{4em}{1987} 
		&  1.185 & 0.832  & 0.706
		\\
		
		& 			\multirow{1}{4em}{1988} 
		&  1.659 & 1.533 & 0.814
		\\
		& 			\multirow{1}{4em}{1989} 
		&  1.183 & 0.895 & 0.717
	\\
		\hline
		\\
	\end{tabularx}
		\label{fig:table5}

\end{table*}

\subsection{Result: Measuring Tail Risk using VaR}
Table \ref{fig:table6}, \ref{fig:table7}, \ref{fig:table8} and \ref{fig:table9}
provides the estimates of VaRs at the $90^{\text{th}}$ percent level and the $95^{\text{th}}$ percent level, along with the asymptotic 95\% CIs of the VaRs. We labelled the CIs that does not cover the empirical VaRs with red color. Similar to what we described in section 3.3, the accumulation tests (in Tables \ref{fig:table7}, \ref{fig:table8} and \ref{fig:table9}) generally produced closer estimates to the empirical VaRs compared to the previous graphical methods (in Table \ref{fig:table6}) over all the years. In addition, notice when using the previous graphical methods, all the asymptotic 95\% CIs for the VaRs did not cover the corresponding empirical estimates. This implies the graphical methods generally could not provide satisfactory fittings for the high quantiles. 
\par The 95\% asymptotic CIs for VaRs were computed based on the Delta method described in section 2.6. We labeled a CI in red color if the CI does not cover the corresponding empirical VaR estimate. We noticed that the CIs produced from the ForwardStop selection covered almost all the empirical VaR estimates. This implies the ForwardStop selection has the best ability to model the high quantiles of the claim size distributions in comparison with the SeqStep and HingeExp selections.

\begin{table*}[!htb]
	\centering
	\caption{Value-at-Risks of the Fitted GPD Models using Previous Graphical Methods with Norwegian Fire Insurance Claims (1985-1992)}
		\footnotesize
\begin{threeparttable}
	\begin{tabularx}{\textwidth}{XXXXX}
		\hline
		\multirow{2}{4em}{Year} & \multirow{2}{4em}{Method}
			&\multirow{2}{10em}{VaR(0.90) (95\% CI)}  & \multirow{2}{10em}{VaR(0.95) (95\% CI)}  
		\\
		
		\\
		
		\hline
		
		  \multirow{3}{4em}{1985} & Empirical & 3.50 & 7.15 
		\\

		&GW	& 22.13 \textcolor{red}{(18.92, 25.35)} & 40.26 \textcolor{red}{(33.89, 46.84)}
		\\
		
		&MRL	& 29.46 \textcolor{red}{(25.40, 33.53)} & 52.60 \textcolor{red}{(44.81, 66.39)}
		\\
		\hline
		  \multirow{3}{4em}{1986} & Empirical  & 3.34 & 6.97 
		\\

		&GW	&25.85 \textcolor{red}{(21.84, 29.86)}& 51.12 \textcolor{red}{(42.09, 60.15)}
		\\
		
		&MRL	& 35.10 \textcolor{red}{(30.45, 39.75)} & 63.29 \textcolor{red}{(54.26, 72.33)}
		\\
		\hline
		  \multirow{3}{4em}{1987} & Empirical & 3.52 & 6.04
		\\

		&GW	&17.19 \textcolor{red}{(15.46, 18.93)} & 25.89 \textcolor{red}{(23.02, 28.75)}
		\\
		
		&MRL	& 14.72 \textcolor{red}{(16.09, 20.52)} & 24.04 \textcolor{red}{(20.88, 27.21)}
		\\
		\hline
		  \multirow{3}{4em}{1988} & Empirical &  4.55& 7.72 
		\\

		&GW	& 23.16 \textcolor{red}{(20.46, 25.86)}& 40.52 \textcolor{red}{(35.35, 45.69)}
		\\
		
		&MRL	& 32.69 \textcolor{red}{(29.06, 36.32)} & 56.11 \textcolor{red}{(49.35, 62.86)}
		\\
		\hline
				  \multirow{3}{4em}{1989} & Empirical & 3.86 & 6.19  
		\\

		&GW	& 18.31 \textcolor{red}{(16.10, 20.52)} & 30.71 \textcolor{red}{(26.60, 34.82)}
		\\
		
		&MRL	& 51.60 \textcolor{red}{(46.25, 56.96)} & 75.77 \textcolor{red}{(67.40, 84.13)}
		\\
		\hline

		\hline
		\\
	\end{tabularx}
		\label{fig:table6}
\begin{tablenotes}
\item[*] Red color indicates the CI does not cover the corresponding empirical VaR estimate.
\end{tablenotes}
\end{threeparttable}
\end{table*}
\begin{table*}[!htb]
	\centering
	\caption{Value-at-Risks of the Fitted GPD Models using ForwardStop Selection with Norwegian Fire Insurance Claims (1985-1992)}
		\footnotesize
\begin{threeparttable}
	\begin{tabularx}{\textwidth}{XXXXX}
		\hline
		\multirow{2}{4em}{Year} & \multirow{2}{4em}{$\alpha$}
			&\multirow{2}{10em}{VaR(0.90) (95\% CI)}  & \multirow{2}{10em}{VaR(0.95) (95\% CI)}  
		\\
		
		\\
		
		\hline
		
		  \multirow{3}{4em}{1985} & Empirical  & 3.50 & 7.15 
		\\

		&0.01	& 4.07 (2.34, 5.81) & 7.08 (2.93, 11.23) 
		\\
		
		&0.05	&4.18 (1.92, 6.44) & 7.43 (1.77, 13.10)
		\\
		\hline
		  \multirow{3}{4em}{1986} & Empirical  & 3.34 & 6.97 
		\\

		&0.01	& 4.24 (2.31, 6.18) & 7.32 (2.63, 12.13)
		\\
		
		&0.05	&6.71 (3.29, 10.12)  & 12.75 (3.41, 22.08)
		\\
		\hline
		  \multirow{3}{4em}{1987} & Empirical  & 3.52 & 6.04 
		\\

		&0.01	& 4.19 \textcolor{red}{(3.54, 4.84)} & 6.48 (5.26, 7.70)
		\\
		
		&0.05	&5.15 \textcolor{red}{(4.22, 6.08)}   & 8.26 \textcolor{red}{(6.33, 10.19)}
		\\
		\hline
		  \multirow{3}{4em}{1988} & Empirical  &  4.55& 7.72 
		\\

		&0.01	& 5.61 (4.30, 6.91)  & 9.73  (6.76, 12.70)
		\\
		
		&0.05	& 6.85 \textcolor{red}{(5.36, 8.34)} & 12.09 \textcolor{red}{(8.67, 15.52)}
		\\
		\hline
				  \multirow{3}{4em}{1989} & Empirical  & 3.86 & 6.19  
		\\

		&0.01	& 4.16 (3.45, 4.86) & 6.55 (5.22, 7.88)
		\\
		
		&0.05	&4.20 (3.46, 4.94)  & 6.67 (5.24, 8.09)
		\\
		\hline

		\hline
		\\
	\end{tabularx}
		\label{fig:table7}
\begin{tablenotes}
\item[*] Red color indicates the CI does not cover the corresponding empirical VaR estimate.
\end{tablenotes}
\end{threeparttable}
\end{table*}

\begin{table*}[!htb]
	\centering
	\caption{Value-at-Risks of the Fitted GPD Models using SeqStep (C=2) Selection with Norwegian Fire Insurance Claims (1985-1992)}
		\footnotesize
\begin{threeparttable}
	\begin{tabularx}{\textwidth}{XXXXX}
		\hline
		\multirow{2}{4em}{Year} & \multirow{2}{4em}{$\alpha$}
			&\multirow{2}{10em}{VaR(0.90)}  & \multirow{2}{10em}{VaR(0.95)}   
		\\
		
		\\
		
		\hline
		
		  \multirow{3}{4em}{1985} & Empirical & 3.50 & 7.15 
		\\
		
		&0.01	& 5.88 (3.50, 8.26) & 10.71 (4.68, 16.74)  
		\\
		
		&0.05	&8.17 \textcolor{red}{(5.63, 10.71)} & 15.26 \textcolor{red}{(8.84, 21.68)}
		\\
		
		\hline
		  \multirow{3}{4em}{1986} & Empirical  & 3.34 & 6.97 
		\\

		&0.01	& 6.71 (3.29, 10.12) & 12.75 (3.41, 22.08)
		\\
		
		&0.05	&7.32 \textcolor{red}{(3.65, 10.98)} &14.14 (3.96, 24.32)
		\\
		\hline
		  \multirow{3}{4em}{1987} & Empirical  & 3.52 & 6.04 
		\\

		&0.01	& 5.61 \textcolor{red}{(4.63, 6.59)} & 9.04 \textcolor{red}{(7.00, 11.08)}
		\\
		
		&0.05	&5.67 \textcolor{red}{(4.67, 6.67)} &9.16 \textcolor{red}{(7.08, 11.24)}
		\\
		\hline
		  \multirow{3}{4em}{1988} & Empirical  &  4.55& 7.72 
		\\

		&0.01	& 8.43 \textcolor{red}{(6.37, 10.49)} & 15.62 \textcolor{red}{(10.51, 20.73)}
		\\
		
		&0.05	& 11.46 \textcolor{red}{(9.70, 13.21)} &20.64 \textcolor{red}{(16.84, 24.43)}
		\\
		\hline
				  \multirow{3}{4em}{1989} & Empirical  & 3.86 & 6.19  
		\\

		&0.01	& 6.26 \textcolor{red}{(5.17, 7.36)} & 10.25 \textcolor{red}{(7.98, 12.53)}
		\\
		
		&0.05	&6.34 \textcolor{red}{(5.15, 7.55)} &10.52 \textcolor{red}{(7.95, 13.08)}
		\\
		\hline

		\hline
		\\
	\end{tabularx}
		\label{fig:table8}
\begin{tablenotes}
\item[*] Red color indicates the CI does not cover the corresponding empirical VaR estimate.
\end{tablenotes}
\end{threeparttable}
\end{table*}

\begin{table*}[!htb]
	\centering
	\caption{Value-at-Risks of the Fitted GPD Models using HingeExp (C=2) Selection with Norwegian Fire Insurance Claims (1985-1992)}
		\footnotesize
\begin{threeparttable}
	\begin{tabularx}{\textwidth}{XXXXX}
		\hline
		\multirow{2}{4em}{Year} & \multirow{2}{4em}{$\alpha$}
			&\multirow{2}{10em}{VaR(0.90)}  & \multirow{2}{10em}{VaR(0.95)}  
		\\
		
		\\
		
		\hline
		
		  \multirow{3}{4em}{1985} & Empirical & 3.50 & 7.15 
		\\

		&0.01	& 8.32 \textcolor{red}{(5.74, 10.89)} & 15.57 \textcolor{red}{(9.04, 22.10)}
		\\
		
		&0.05	&9.47 \textcolor{red}{(6.10, 12.85)}&18.51 \textcolor{red}{(9.41, 27.61)}
		\\
		\hline
		  \multirow{3}{4em}{1986} & Empirical  & 3.34 & 6.97 
		\\

		&0.01	&7.32 (3.65, 10.98)&14.14 (3.96, 24.32)
		\\
		
		&0.05	&47.32 \textcolor{red}{(40.25, 54.40)}&100.38\textcolor{red}{(83.36,117.39)}
		\\
		\hline
		  \multirow{3}{4em}{1987} & Empirical & 3.52 & 6.04
		\\

		&0.01	&5.67 \textcolor{red}{(4.67, 6.67)} &9.16 \textcolor{red}{(7.08, 11.24)}
		\\
		
		&0.05	&6.00 \textcolor{red}{(4.91, 7.08)}&9.78 \textcolor{red}{(7.48, 12.07)}
		\\
		\hline
		  \multirow{3}{4em}{1988} & Empirical &  4.55& 7.72 
		\\

		&0.01	& 11.46 \textcolor{red}{(9.70, 13.21)}&20.64 \textcolor{red}{(16.84, 24.43)}
		\\
		
		&0.05	&12.05 \textcolor{red}{(10.35, 13.74)} &21.35 \textcolor{red}{(17.84, 24.86)}
		\\
		\hline
				  \multirow{3}{4em}{1989} & Empirical & 3.86 & 6.19  
		\\

		&0.01	& 6.26 \textcolor{red}{(5.17, 7.36)} & 10.25 \textcolor{red}{(7.98, 12.53)}
		\\
		
		&0.05	&6.44 \textcolor{red}{(5.27, 7.61)} &10.63 \textcolor{red}{(8.17, 13.09)}
		\\
		\hline

		\hline
		\\
	\end{tabularx}
		\label{fig:table9}
\begin{tablenotes}
\item[*] Red color indicates the CI does not cover the corresponding empirical VaR estimate.
\end{tablenotes}
\end{threeparttable}
\end{table*}

\section{Conclusion}

In this article, three different accumulation tests have been used to generate the GPD models for claim size distributions and VaR estimates over the Norwegian Fire Insurance Data. Among the accumulation tests, the ForwardStop utilizes a smooth logarithmic function as the accumulation function while the SeqStep and the HingeExp employ a discrete step function with a pre-specified parameter $C$ as the accumulation function. We also used the previous graphical methods including GW-Plot and MRL-Plot to generate the models for comparison purposes. Among all the models, the ForwardStop selection demonstrates the best performance as it produces the closest fits to the empirical CDFs and the closest estimates to the empirical VaRs at the 90\% and 95\% level. The SeqStep and the HingeExp also performed better in comparison with previous graphical methods. Therefore, regarding the Norwegian Fire Insurance Data, the threshold selection methods based on the accumulation tests have better ability to capture the features of the claim size distributions compared to the previous widely-used methods.   
\par The threshold selection based on the accumulation tests is attractive due to their good performance fitting the Norwegian Fire Insurance Data. However, one needs to take precautions in terms of the following: the quality of the modeling for the accumulation tests are affected by the chosen FDP ($\alpha$). For instance, the ForwardStop selection with $\alpha = 0.01$ generally overperformed the same procedure with $\alpha = 0.05$. Hence, before using the accumulation tests, a reasonable selection on $\alpha$ is necessary in order to obtain a better model, especially for modeling high quantiles. In addition, both SeqStep and HingeExp use a specific parameter $C$ that needs to be specified before the selection procedure. We did not test the effect of $C$ on the estimation of GPD modeling in this article. However, it definitely has an impact on the selection of thresholds since different $C$ will lead to different choices of the threshold $\mu$, even with the same dataset. 
\par Finally, we conclude that the accumulation tests have great ability to choose the thresholds in the GPD models for the Norwegian Fire Insurance Data. More extended uses and improvements for such method in the GPD modeling are warranted, especially producing tail risk measures other than VaR (e.g. tail conditional mean) by using the thresholds chosen by the accumulation tests.

\bibliography{seq_gpd_ref,composite_insurance}

\end{document}